\documentclass[letterpaper,twocolumn,nofootinbib, nobibnotes,aps,prd]{revtex4}
\pdfoutput=1

\usepackage{graphicx}
\usepackage{amsmath,amsthm,amssymb}
\def\gr{$\gamma$-ray}

\newcommand{\be}{\begin{equation}}
\newcommand{\ee}{\end{equation}}
\newcommand{\ba}{\begin{eqnarray}}
\newcommand{\ea}{\end{eqnarray}}

\begin{document}

\title{Galactic and extragalactic contributions to the astrophysical muon neutrino signal}
 \author{Andrii Neronov$^{1}$}
 \author{Dmitry Semikoz$^{2,3}$}
\affiliation{$^{1}$ISDC, Astronomy Department, University of Geneva, Ch.~d'Ecogia 16, Versoix 1290, Switzerland}
\affiliation{$^{2}$APC, Universite Paris Diderot, CNRS/IN2P3, CEA/IRFU, Observatoire de Paris, Sorbonne Paris Cite,
119 75205 Paris, France}
\affiliation{$^{3}$ National Research Nuclear University “MEPHI” (Moscow Engineering Physics Institute), Kashirskoe highway 31, Moscow, 115409, Russia}
\begin{abstract}
Spectral and anisotropy properties of  IceCube astrophysical neutrino signal reveal an evidence for  a significant Galactic contribution to the neutrino flux in Southern hemisphere. We check if the Galactic contribution is detectable in the astrophysical muon neutrino flux observed from a low positive declinations region of the Northern sky.  Estimating the Galactic neutrino flux in this part of the sky from  \gr\ and Southern sky neutrino data, we find that the Northern sky astrophysical muon neutrino signal shows an excess over the Galactic flux. This points to the presence of an additional hard spectrum (extragalactic or large scale Galactic halo) component of astrophysical neutrino flux. We show that the Galactic flux component should still be detectable in the muon neutrino data in a decade long IceCube exposure.
\end{abstract}

\maketitle

\section{Introduction}

\begin{figure}
\includegraphics[width=\linewidth]{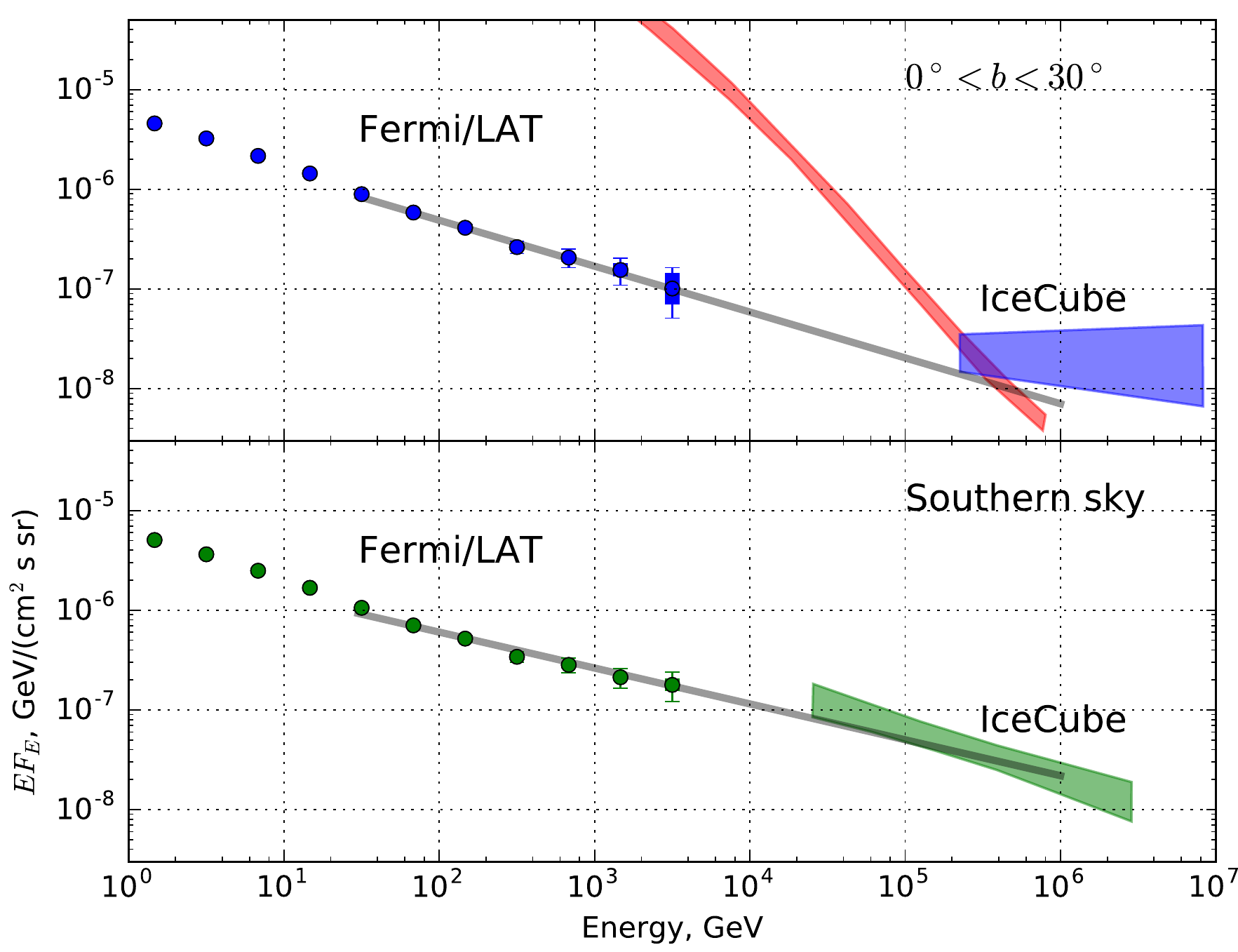}
\caption{Spectra of neutrino and \gr\ emission from Southern sky (bottom panel) and from a low declinations ($0^\circ<DEC<30^\circ$) strip (top panel) Data points show the \gr\ spectrum, shaded regions are fro the neutrino spectra. Red shaded band in the bottom panel shows the atmospheric neutrino background.}
\label{fig:south}
\end{figure}

IceCube collaboration has recently reported the detection of astrophysical neutrino signal in the energy range from 30 TeV to 2 PeV \cite{IceCube_1yr,IceCube_PeV,IceCube_3yr,IceCube_2yr,IceCube_combined_2015,IceCube_muonnu_2015}. The signal is observed in two different modes: the high-energy starting neutrino  events (HESE) produced by neutrinos of all flavours interacting in the detector and the muon track events produced by muon neutrinos interacting in the vicinity of the detector. The HESE spectrum is compatible with a powerlaw   \cite{IceCube_combined_2015} with the slope $dN_\nu/dE\propto E^{-\Gamma}$, $\Gamma=2.5\pm 0.1$ above 30~TeV (Fig. \ref{fig:south}, bottom panel).  This signal is mostly collected from the Southern sky exposed to the inner part of the Milky Way galaxy. The muon neutrino signal reveals a harder powerlaw \cite{IceCube_muonnu_2015} with the slope $\Gamma=-1.91\pm 0.2$ in the energy range above 250~TeV (Fig. \ref{fig:south}, top panel). It is dominated by a signal from a low declinatons strip along the equatorial plane. 

The origin of  the astrophysical neutrino signal is not clear. The overall sky distribution of the three-year HESE sample is consistent with an isotropic distribution \cite{IceCube_3yr}. This could be considered as an argument in favour of extragalactic origin of the signal \cite{stecker91,waxmanbachall,murase14,tkachev_agn,Giacinti:2015pya,murase_ahlers}. At the same time, evidence (at $3\sigma$ level) for a  large scale anisotropy of HESE arrival directions correlated with the orientation of the Galactic plane and Galactic poles points to the presence of Galactic contribution to the flux  \cite{three_sigma}. This evidence is supported by the consistency of the \gr\ and neutrino spectra in the sky region from which the HESE events are collected \cite{Neronov:2013lza,Neronov:2014uma}. This possibility would imply a significant revision of the modelling of propagation of cosmic rays in the interstellar medium \cite{kachelriess14,joshi13,ahlers15}.

Variation of the spectral properties of neutrino signal across the sky could potentially also point to the presence of Galactic flux contribution, because the extragalactic flux is expected to be isotropic and have the same spectrum in all directions.  An unambiguous conclusion on the variability of the spectrum across the sky is however hampered by slight difference of the range of energies of neutrinos which contribute to the HESE and muon astrophysical neutrino signals. It is in principle possible that the spectrum of isotropic neutrino flux experiences a break at  several hundred TeV energy (e.g. due to the presence of two different extragalactic source populations with different spectral characteristics).    

Variations of neutrino spectrum across the sky could be intrinsic to the Galactic  flux component (different cosmic ray populations produce different neutrino spectra in different regions of the Galaxy).  Otherwise, the large scale variations of the spectrum could be due to the presence of both Galactic and extragalactic contributions to the flux. The Galactic flux component might dominate the flux from the inner Galaxy and be sub-dominant in the outer Galaxy and / or at high Galactic latitudes. 

In what follows we explore these two possibilities by complementing the IceCube neutrino data with the lower energy \gr\ data collected by Fermi Large Area Telescope (LAT) \citep{atwood09}. We estimate the normalisation and slope of the Galactic component of neutrino spectrum in different sky regions from the \gr\ data. We show  the HESE and muon neutrino signals are different in the sense that the former is consistent with the high-energy extrapolation of the Galactic \gr\ spectrum, while the latter shows an excess with respect to the estimated Galactic flux. The Northern sky excess should be due to the presence of an additional hard spectrum component  (extragalactic?). We re-iterate on the conclusion of Ref. \cite{tchernin13a} that detection of the Galactic component of the flux should still be possible also in the muon neutrino signal in a decade time scale exposure, even in the presence of the additional unrelated hard spectrum component of the flux.

\section{Galactic contribution to the astrophysical muon neutrino flux}

Strong atmospheric neutrino background prevents detection of diffuse astrophysical muon neutrino signal at the energies below several hundred TeV (see Fig. \ref{fig:south}, bottom panel). The Earth is only partially transparent to neutrinos neutrinos with 0.1-10~PeV energy. This explains the fact that the astrophysical muon neutrino signal is detected in a low declinations strip along the equatorial plane. 

The flux of the astrophysical signal is approximately at the level of the atmospheric neutrino flux at $E\simeq 200$~TeV and starts to dominate over the atmospheric background above approximately 400~TeV. The $E>400$~TeV event sample reported in the Ref. \cite{IceCube_muonnu_2015} has 7 events with only one estimated background. These events are distributed in the declination range $0^\circ<DEC<30^\circ$. 

To estimate the Galactic component of the muon neutrino flux in this sky region, we have extracted the \gr\ spectrum of  the strip using the data of Fermi/LAT telescope. We have processed the Pass 8 \gr\ data using Fermi Science Tools version v10r0p5. The data were filtered using the {\it gtselect} -- {\it gtmktime} chain following the recommendations of the Fermi Science Support Centre team \footnote{http://fermi.gsfc.nasa.gov/ssc/data/analysis/}. The spectrum of the $0^\circ<DEC<30^\circ$ strip was extracted using the "aperture photometry" method after calculation of exposure on a $20^\circ$ step grid of points within the strip using the {\it gtexposure} tool. We have used the "CLEAN" event selection which has the minimal residual cosmic ray content. To get the spectrum of diffuse (Galactic plus isotropic) emission we have excluded from the data set circles of the radius $1^\circ$ around identified sources from the four-year Fermi catalog \cite{fermi_catalog}. We did not subtract the unidentified sources, because some of those sources could be part of the Galactic  diffuse emission, found at the locations of denser clumps of the interstellar medium. The resulting spectrum of the $0^\circ<DEC<30^\circ$ strip is shown in Fig. \ref{fig:south}. For completeness, we have also extracted the spectrum of the Southern sky ($DEC<0^\circ$) using the same approach. This spectrum is also shown in Fig. \ref{fig:south}. 

\begin{figure}
\includegraphics[width=\linewidth]{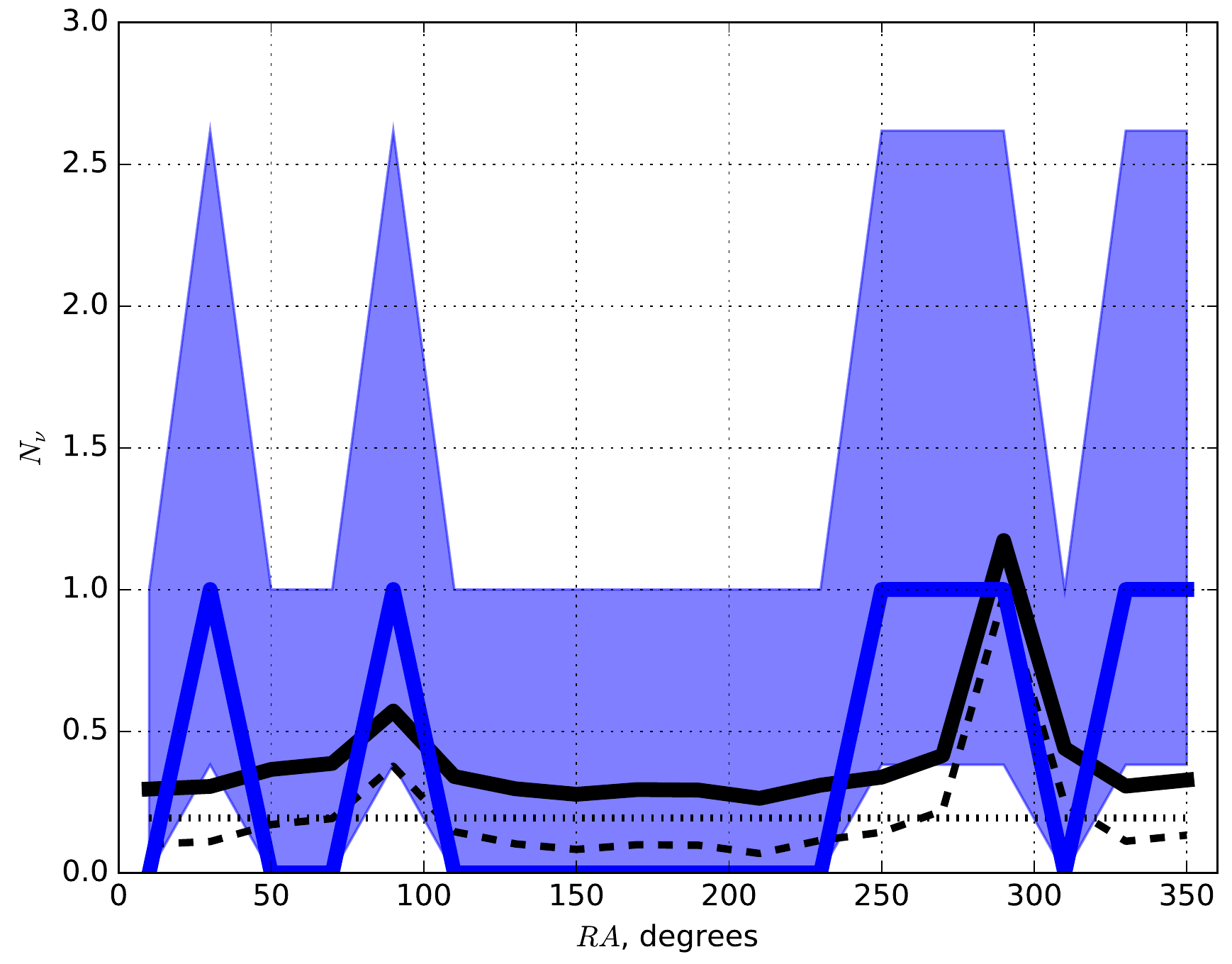}
\caption{Distribution of arrival direction of muon neutrino events with estimated energies above 400~TeV in the Right Accention (blue line and blue shaded uncertainty band. Thick black solid curve shows a two component model composed of isotropic  (dotted line) and  Galactic (dashed line) contributions with equal overall flux.  }
\label{fig:profile}
\end{figure}

Fig. \ref{fig:south} also shows a comparison of the \gr\ and neutrino spectra of the $0^\circ<DEC<30^\circ$ and $DEC<0^\circ$ parts of the sky. One could observe an essential difference of the "multi-messenger" spectra of these sky regions. The Southern sky multi-messenger spectrum shows a smooth match between the \gr\ and neutrino spectra. The normalisation and slope of high-energy extrapolation of the \gr\ spectrum from $E\lesssim 5$~TeV energy range to $E>30$~TeV range are consistent with those of the neutrino spectrum above 30~TeV. As it is noticed in the Ref. \cite{Neronov:2014uma}, this points to the common Galactic origin of the  \gr\ and neutrino emission in this part of the sky. 

To the contrary, the high-energy extrapolation of the \gr\ spectrum of the $0^\circ<DEC<30^\circ$ strip is not consistent with the astrophysical muon neutrino spectrum observed in this part of the sky. One could see that the best fit powerlaw model of the \gr\ spectrum under-predicts the neutrino flux at $\simeq 400$~TeV energy by a factor of 2. Moreover, the best fit slope of the \gr\ spectrum, $\Gamma_\gamma=2.46$ is softer than the slope of neutrino spectrum ($\Gamma_\nu=1.91\pm 0.2$ \cite{IceCube_muonnu_2015}). Thus, the Northern sky astrophysical neutrino flux shows a clear excess above the Galactic flux calculated under an assumption of a powerlaw distribution of Galactic cosmic rays.  This hard spectrum excess could be of extragalactic origin \cite{stecker91,mannheim92,neronov02} or it could originate from a larger scale Galactic halo trapping the cosmic rays escaping from the Galactic Disk \cite{taylor14}.  

Fig. \ref{fig:profile} shows the distribution of muon neutrino events with estimated energies above 400~TeV in Right Accention. There are 7 events in the IceCube data set discussed in Ref.  \cite{IceCube_muonnu_2015}. This event set is almost atmospheric background free, only one out of seven events is expected to come from the background (to the contrary, half of the lower energy events from Ref. \cite{IceCube_muonnu_2015} are background). Model curves in Fig. \ref{fig:profile} show a comparison of the observed distribution of events with the model based on the best-guess estimate of the Galactic component of the flux  (half of neutrino flux at 400~TeV).   One could see that the model is consistent with the data. 

The Galactic component should reveal itself at the first place in the direction where the observation strip $0^\circ<DEC<30^\circ$ crosses the inner Galactic Plane. A weaker and broader Galactic excess is expected in the direction of intersection of the observation strip with the outer Galactic Plane in the $RA\simeq 100^\circ$ bin. With the two-year exposure, the inner Galaxy intersection region is expected to produce only one neutrino event. Occasionally, there is one neutrino event with energy $E>400$~GeV  in the direction of the inner Galactic Plane in the IceCube data set. 

Obviously, larger exposure time is needed for a significant detection of the excess in the direction of the inner Galactic Plane excess in the muon neutrino data set. Taking into account the analysis of Ref.  \cite{IceCube_muonnu_2015} is based on a two-year exposure, one could estimate that a two-decade exposure would result in $\sim 10$ events in the direction of the inner Galactic Plane and for an overall $\sim 35$ event statistics of the Galactic flux integrated over the $0^\circ<DEC<30^\circ$ strip.  In the absence of additional hard spectrum component of astrophysical neutrino flux, the event statistics would be sufficient for a discovery of the Galactic flux, because the flux would be detected on top of residual atmospheric neutrino background (with comparable statistics, homogeneously distributed over Right Accention). 
Presence of the additional hard component makes the detection more difficult. Only the difference in the $RA$ distribution of Galactic and isotropic signal could be used to distinguish the Galactic component from the isotropic component (not the overall signal statistics). The excess of $\sim 10$ events over the expected $\sim 3$ events of isotropic component in the RA bin containing the intersection with the inner Galactic Plane would amount only to a $\sim 3\sigma$ evidence for the presence of the Galactic flux. This conclusion is in agreement with the result of analysis of Ref.   \cite{tchernin13a} which noticed that detection of Galactic neutrino flux in Northern hemisphere would require decade-scale exposures of IceCube. 

\begin{figure}
\includegraphics[width=\linewidth]{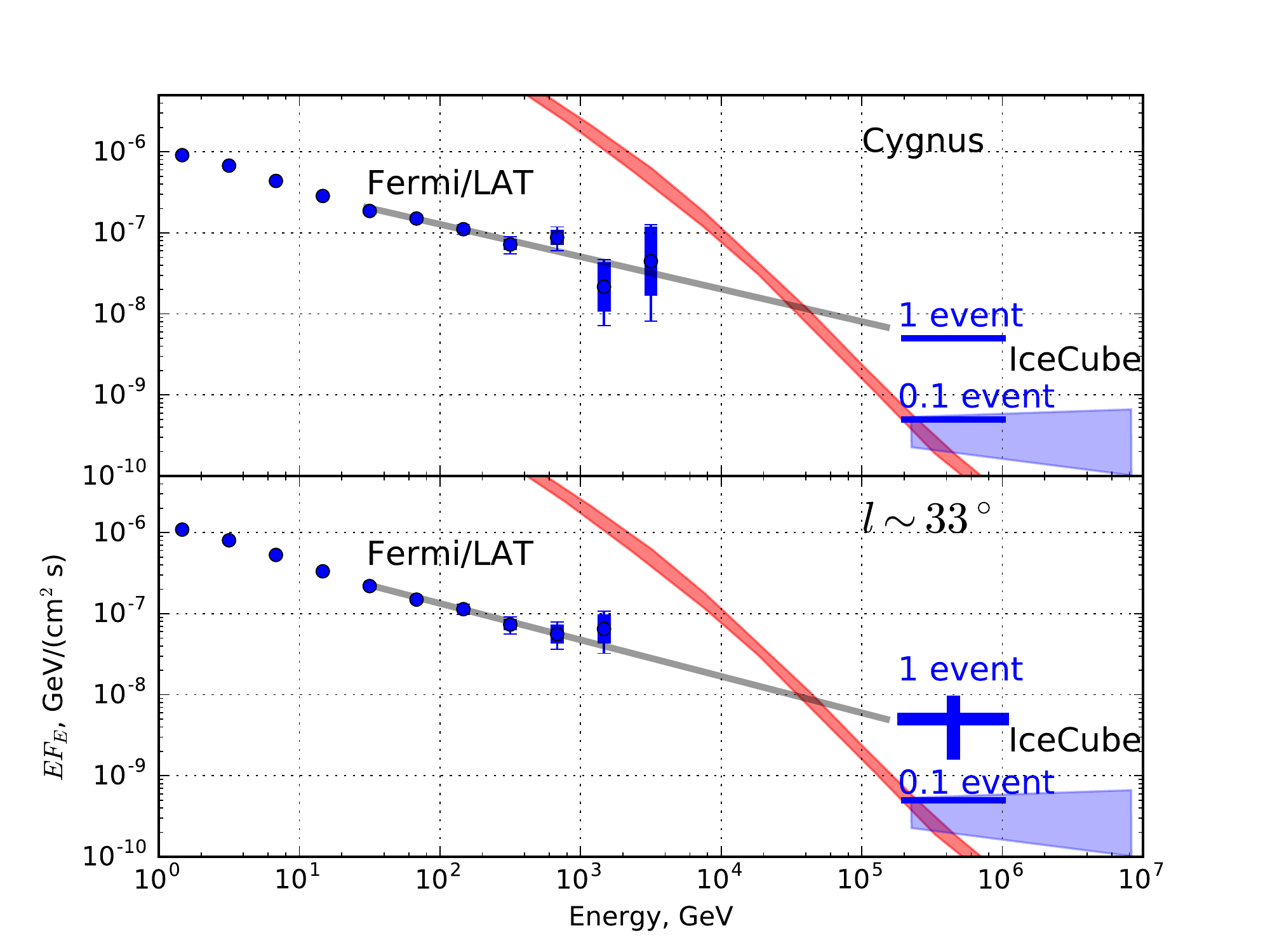}
\caption{\gr\ and neutrino spectra of two strongest extended excesses of Galactic emission in the Northern sky \cite{tchernin13a}. Notations are the same as in Fig. \ref{fig:south}. The shaded areas show  the expected level of isotropic neutrino flux. Horizontal tics mark the flux levels corresponding to 0.1 and 1 events in two-year IceCube exposure. Vertical error bar for neutrino data in the bottom panel shows the uncertainty of the flux estimate based on one detected event from the HESS J1857+026 region. }
\label{fig:cygnus}
\end{figure}

Ref.  \cite{tchernin13a} has reached a conclusion that the best strategy for identification of Galactic component of neutrino flux in the Northern hemisphere is to search for localised extended emission regions with excess cosmic ray densities. The two regions with the strongest excess of Galactic neutrino flux in the Northern sky are expected to be the region of HESS J1857+026 and Cygnus X region. The extended excess search strategy helps to reduce the unrelated homogeneously distributed background and detect weaker Galactic signal in the Northern sky. The same conclusion is valid also in the presence of the hard spectrum isotropic astrophysical neutrino flux, which serves as an additional background component for the search of the Galactic emission (together with the atmospheric neutrino flux). Fig. \ref{fig:cygnus} shows the expected Galactic neutrino flux from circles of the radius $4^\circ$ in the  directions of Cygnus X and HESS J1857+026 regions (centred at coordinates defined in Ref. \cite{tchernin13a}) calculated based on extrapolation of the \gr\ signal, using the same approach as for the $0^\circ<DEC<30^\circ$ strip.  One could see that contrary to the Galactic flux averaged over the observation region, the Galactic flux estimated from the \gr\ data is predicted to dominate over the hard spectrum component in these particular directions. 

The statistics of the two-year data reported in Ref. \cite{IceCube_muonnu_2015} is not yet sufficient for detection of the two extended excesses. Both sources are expected to produce approximately one detectable neutrino with this exposure. In fact, one neutrino is detected from the direction of HESS J1857+026 region. Estimate of the neutrino flux corresponding to one event, shown in the botom panel of Fig. \ref{fig:cygnus} is consistent with the high-energy extrapolation of Fermi/LAT flux. Taking into account that the source flux exceeds the hard spectrum atropysical background by a factor of $\simeq 10$, a $5\sigma$ detection of the HESS J1857+026 region excess would require 10 events (with 1 background). Accumulation of such statistics requires a two decade long IceCube exposure.

\section{Conclusions}

To summarise,  we have shown that the astrophysical muon neutrino signal is consistent with the estimate of the Galactic neutrino flux suggested by the anisotropy and spectral properties of the HESE neutrino data set and by the Fermi/LAT \gr\ data.  At the same time, the astrophysical muon neutrino signal reveals the presence of an additional hard spectrum excess on top of the estimated Galactic flux. This suggests  a self consistent  interpretation of the astrophysical neutrino signal should be based on a two component model. The softer spectrum Galactic component provides a dominant contribution to the flux in the Southern sky exposed to the inner Galaxy.  The harder component visible in the muon neutrino data appears in the Northern hemisphere exposed to the outer Galaxy. The hard spectrum component could be either extragalactic or originate from a large scale Galactic halo. 

Presence of the additional hard spectrum component makes the detection of Galactic neutrino emission in $E>100$~TeV band difficult, because it creates background for such a detection  (in addition to the residual atmospheric neutrino background). We have shown that the best strategy for identification of Galactic component in the Northern sky observations in the muon neutrino channel is to look for isolated extended excesses in the directions of overdensities of cosmic rays and/or interstellar medium. Detection of the two previously identified strongest extended excesses in the Cygnus and HESS J1857+026 regions above 100~TeV would require an IceCube exposure longer than decade. The detection is possible within a much shorter exposure with IceCube Generation 2 \cite{IceCube_gen2}. 
 
The best possibility for the study of the Galactic component in the muon neutrino channel is through the use of Northern hemisphere neutrino telescopes, such as ARCA detector of km3net \cite{km3net}. The results of analysis of the data of ANTARES  neutrino telescope show that already the sensitivity of this telescope is marginally reaching the expected flux level of the Galactic component (provided that it constitutes 100\% of the IceCube signal) \cite{Antares_ICRC15}. This shows that a detailed study of the spectral and morphological properties of the Galactic signal will be possible with ARCA, with angular resolution superior to that of the HESE event sample of IceCube and starting from the TeV energy band directly overlapping with Fermi/LAT and CTA \gr\  telescope energy bands. 
\vskip-1cm

\section*{Acknowledgement}

We would like to thank C.Wiebusch for useful discussions.

\bibliography{two_components}

\end{document}